\documentclass{epsconf}
\usepackage{graphicx}
\usepackage{wrapfig}
\usepackage{amsmath}

\title{Statistical theory of intermittency in a multi-scale model of MHD and micro-turbulence}
\author{\underline{Johan Anderson}$^1$, Eun-jin Kim$^1$}
\institute{$^1$ Applied Mathematics, University of Sheffield, Sheffield, UK}

\begin{document}
\maketitle
Traditionally the effects of MHD instabilities and micro-instabilities on plasma 
confinement are investigated separately. However, these two instabilities often 
occur simultaneously, with the overlap of the dynamics on a broad range of spatial 
scales. It is thus vital to incorporate these instabilities consistently
by a proper multi-scale modeling. Furthermore,  
there has been an overwhelming evidence that the overall transport of heat and 
particles is significantly influenced by intermittency (or bursty events) 
caused by coherent structures. A crucial question in plasma confinement is 
thus the prediction of the probability distribution functions (PDFs) of the transport 
due to these structures and of their formation.

In this paper, we investigate intermittent transport in
a multi-scale model by consistently incorporating  both tearing instabilities 
and micro-instability due to pressure gradient~\cite{a1}. 
We first present an exact nonlinear solution in the form of a coherent structure 
(modon or bipolar vortex soliton). We then present a first analytical result of 
intermittency in our multi scale model by utilizing a novel non-perturbative method.
Specifically, we compute the PDF tails of momentum flux and heat flux, 
by assuming that a short-lived modon is a coherent structure responsible 
for bursty and intermittent events, contributing to the PDF tails.

The governing equations for modeling the MHD and drift wave turbulence~\cite{a1} are as follows,
\begin{eqnarray}
\frac{\partial}{\partial t} \nabla_{\perp}^2 \phi + [\phi, \nabla_{\perp}^2 \phi] & = & [\psi, \nabla_{\perp}^2 \psi] - \kappa_1 \frac{\partial p}{\partial y} - \nu \nabla_{\perp}^4 \phi + f_1, \\
\frac{\partial p}{\partial t} + [\phi, p] & = & -v_{\star} \left( (1-\kappa_2)\frac{\partial \phi}{\partial y} + \kappa_2 \frac{\partial p}{\partial y}\right) + C^2 [\psi, \nabla_{\perp}^2 \psi] + \chi_i \nabla_{\perp}^2 p + f_2, \\
\frac{\partial \psi}{\partial t} + [\phi - p, \psi] & = & - v_{\star} \frac{\partial \psi}{\partial y} + \eta \nabla_{\perp}^2 \psi + f_3.
\end{eqnarray}
The model is a reduced MHD model that includes tearing modes and interchange modes and consists of a set of coupled equations for the electrostatic potential $\phi$, the pressure $p$ and the magnetic flux $\psi$. Eq. (1) is the combined electron and ion momentum balance equations where $\nu$ is the viscosity and Eq. (2) represents the energy conservation (neglecting parallel ion dynamics) with $\chi_i$ as the diffusivity. Eq. (3) is the Ohm's law with the resistivity $\eta$. The magnetic field is assumed to be dominated by a strong component in the $z$-direction. Here the constants are $v_{\star} = \frac{2 \Omega_i \tau_A L_p}{\beta L_{\perp}}$, $\beta = \frac{p_0}{B^2/(2 \mu_0)}$, $L_p$ is the pressure gradient, $L_{\perp}$ is related to island width, $\Omega = \frac{eB}{m_i}$ is the ion cyclotron frequency and $\tau_A$ is the Alfv\`{e}n time. Eqs. (1)-(3) are normalized by using the Alfv\`{e}n speed $v_A = L_{\perp}/\tau_A$. The interchange mode is controlled by curvature and pressure gradients through the parameters $\kappa_1 = 2 \Omega_i \tau_A \frac{L_{\perp}}{R_0}$ and $\kappa_2 = \frac{10 L_P}{3 R_0}$ ($R_0$ is the major radius), respectively. The tearing mode is determined by the parameter $C = \frac{5 T_e}{3L_{\perp}^2} \Omega_i^2 m_i$. The dynamics of magnetic islands are strongly influenced by the magnitude of $C$. The statistics of the forcings ($f_i$, $i = \{1,2,3\}$) is assumed to be Gaussian with a short correlation time modeled by the delta function as
\begin{eqnarray}
\langle f_i(x, t) f_{j}(x^{\prime}, t^{\prime}) \rangle = \delta_{ij} \delta(t-t^{\prime})\kappa_i(x-x^{\prime}),
\end{eqnarray}
and $\langle f_i \rangle = 0$. The angular brackets denote the average over the statistics of the forcing $f_i$.
The delta correlation in time is chosen for the sake of simplicity of the analysis.

For computing PDF tails the basic idea is to associate the bursty event causing intermittency with the creation of a coherent structure (e.g.  blobs, streamers etc.). The creation process is identified with the instanton which is localized in time, existing during the formation of the coherent structure. This idea is embedded in the instanton method, which is a non-perturbative way of calculating the PDF tails. We calculate the PDF tails of local flux $M(\phi(x=x_0))$ that is the second moment of $\phi$ (e.g. momentum flux or heat flux). The PDF tails are expressed in terms of a path-integral using the Gaussian statistics of the forcing~\cite{a21}. The optimum path is then associated with the creation of a short lived coherent structure (among all possible paths) and the action is evaluated using the saddle-point method on the effective action. The probability density function of the flux $R$ can be defined as
\begin{eqnarray}
P(R) =  \langle \delta(M(\phi(x=x_0)) - R) \rangle = \int d \lambda e^{i \lambda R} I_{\lambda},
\end{eqnarray}
where 
\begin{eqnarray}
I_{\lambda} = \langle \exp(-i \lambda M(\phi(x=x_0))) \rangle = \int \mathcal{D} \phi \mathcal{D} \bar{\phi} \mathcal{D} p    \mathcal{D} \bar{p   } \mathcal{D} \psi \mathcal{D} \bar{\psi} e^{-S_{\lambda}}.
\end{eqnarray}
We assume that the instanton solution is of the form $\phi = F(t) \varphi(x,y)$, $p = G(t) \rho(x,y)$ and $\psi = J(t) \chi(x,y)$ where $j = \{ \varphi, \rho, \chi\}$ are the modon solution satisfying $\nabla_{\perp}^2 X_j = \alpha_{1j} X_j + \alpha_{2 j} x$. Furthermore, we assume a linear relation between $\phi$ and $p$. To simplify the computation of the path-integral the saddle-point method will be used where the variational first derivatives in ($\phi, \bar{\phi}, p, \bar{p}, \psi, \bar{\psi}$) of the action $S_{\lambda}$ will be put to zero. The saddle-point solution of the dynamical variable $\phi(x,t)$ of the form $\phi(x,t) = F(t) \varphi$ is called an instanton if $F(t) = 0$ at $t=-\infty$ and $F(t) \neq 0$ at $t=0$.  Note that, the function $\varphi(x)$ here represents the spatial form of the coherent structure. Two limiting cases will be studied in detail; (1) iso-instanton case ($F = G = J$); and (2) the electrostatic limit ($F = G$ and $J=0$). In the iso-instanton case (1) we use forcings in Eqs. (1)-(3) whereas in the electrostatic case we use forcings in Eqs. (1) and (2). Note that the conjugate variables that mediate between the physical variables and the observable flux arise due to the uncertainty in the physical variables of the stochastic forcing.
\begin{figure}
  \includegraphics[height=.3\textheight]{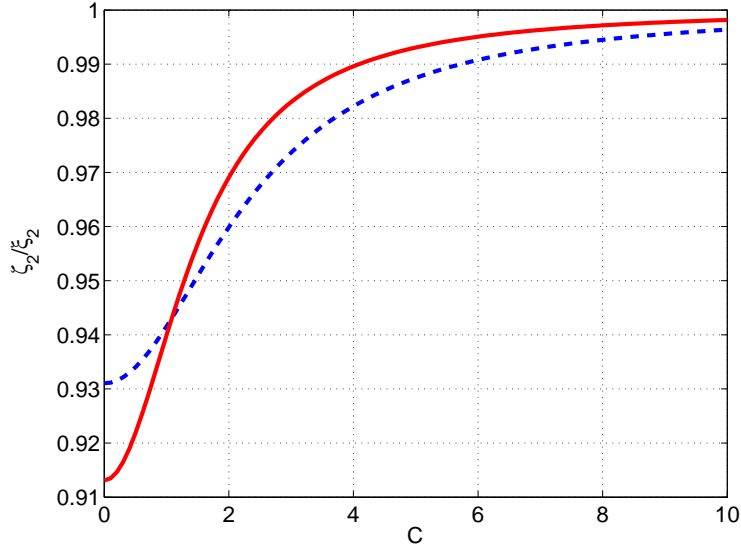}
  \caption{(Color online). The ratio of the coefficients $\xi_2$ (electromagnetic) and $\zeta_2$ (electrostatic) as a function of the parameter $C$, for the other parameters $\kappa_1 = 0.01$ and $\kappa_2 = 0$ (blue line, dashed line) and $\kappa_1 = 0.1$ and $\kappa_2 = 0$ (red line, solid line) are shown.}
\end{figure}

In the iso-instanton case the PDF tails of momentum flux and heat flux are found to be $P(R) \sim \sqrt{R} e^{-\xi_2 R^{3/2}}$ and for particle density $P(R) \sim R^2 e^{-\xi_1 R^3}$. Here the saddle point method with movable maxima~\cite{a32} has been used to compute the integrals in Eqs (5) and (6). Although the power laws multiplying the exponentials have not been included in previous studies, the exponential scaling were obtained previously. The tails of PDF of momentum flux and heat flux are found to be stretched 
exponentials which are broader than a Gaussian. This suggests that rare events of large amplitude due to coherent structure are crucial in transport (similarly to what was found in the previous studies~\cite{a2}, ~\cite{a31}), offering a novel explanation for exponential PDF tails of momentum flux found in recent experiments at CSDX at UCSD~\cite{a4}. These and our previous results highlight the key role of structures on intermittent turbulent transport.

In a similar way the PDF tails in the electrostatic case is computed and the PDF tails of momentum flux and heat flux are found to be $P(R) \sim \sqrt{R} e^{-\zeta_2 R^{3/2}}$ and for particle density flux $P(R) \sim R^2 e^{-\zeta_1 R^3}$, respectively. 

In particular, we will now examine the crucial dependence of the overall amplitude of the PDFs on the relevant physical parameters. We show the effect of magnetic fluctuations on momentum transport by an explicit comparison of the coefficients in the iso-instanton ($\xi_2$) case with the electrostatic case ($\zeta_2$) in Figure 1. It is shown that the values of coefficient in the electromagnetic case is larger than those found in electrostatic case in a scan of the parameter $C$ for two sets of parameters $\kappa_1 = 0.01$ and $\kappa_2 = 0$ (blue line, dashed line) and $\kappa = 0.1$ and $\kappa_2 = 0$ (red line, solid line) (the parameters are similar to those used in Ref~\cite{a1}). Note that by using $\kappa_2 = 0$ the linear diamagnetic effects are suppressed and $v_{\star} = 0$ and that viscosity ($\nu$), diffusivity ($\chi_{\perp}$) and resistivity ($\eta$) are taken small to be $\sim 10^{-4}$. This suggests that the PDF tails in the electromagnetic case is smaller and thus the effects of rare events are smaller in comparison to the electrostatic case. Note that the parameter $C$ determines the temperature response of the tearing modes. Note also that the modon structure is determined by matching the inner and outer solution in the same manner as the drift wave vortex solution in Ref.~\cite{a33}. 

Interestingly, there exists a modon solution where the Maxwell stress is canceled by the Reynolds stress, resulting in a ratio $\zeta_2/\xi_2 \approx 1$ for all values of $C$, $\kappa_1$ and $\kappa_2$. This illustrates the exact cancellation of Maxwell stress by Reynolds stress due to electromagnetic fluctuations. In the present study the effects of pressure gradient are small since the parameter $\kappa_2 = 0$.

We note that treating first the MHD and micro-turbulence as a fully multi-scale model as in Ref.~\cite{a34} and then computing the PDF tails could possibly lead to other exponential dependencies of the PDF tails. This will be investigated in future publications.

{\bf \Large Acknowledgment} \\
This research was supported by the Engineering and Physical Sciences Research Council (EPSRC) EP/D064317/1.

\end{document}